\documentclass[pra,twocolumn, showpacs, showkeys, secnumarabic, aps, amsmath, amssymb, nofootinbib, superscriptaddress, longbibliography, floatfix, table-of-contents, dblfloatfix]{revtex4-2}

\usepackage[pdftex]{graphicx}
\usepackage{mathrsfs}
\usepackage[colorlinks, breaklinks, urlcolor={blue}, linkcolor={red}, citecolor={blue}]{hyperref}
\usepackage{array}
\usepackage{multibib}
\usepackage{amsmath}
\usepackage{type1cm}
\usepackage{lettrine}
\usepackage[english]{babel}
\usepackage{lmodern}
\usepackage{microtype}
\usepackage{booktabs}
\usepackage[T1]{fontenc}
\usepackage[boxed, vlined]{algorithm2e}
\usepackage{caption}
\usepackage{braket}
\usepackage{xcolor}

\begin{document}

\title{Quantum synchronization due to information backflow}

\author{Md.~Manirul Ali}
\email[]{manirul@citchennai.net}
\affiliation{Centre for Quantum Science and Technology, Chennai Institute of Technology, Chennai 600069, India}

\author{Po-Wen Chen}
\email[]{powen@iner.gov.tw}
\affiliation{Physics Division, Institute of Nuclear Energy Research, Longtan, Taoyuan 32546, Taiwan.}

\author{Chandrashekar Radhakrishnan}
\email[]{chandrashekar@physics.iitm.ac.in}
\affiliation{Centre for Quantum Information, Communication and Computing, Indian Institute of Technology Madras, Chennai 600036, India}
\affiliation{Centre for Quantum Science and Technology, Chennai Institute of Technology, Chennai 600069, India}
\homepage[]{https://sites.google.com/view/chandrashekar}

\begin{abstract}
The phase synchronization of a single qubit in a dissipative bath in the absence of driving field is demonstrated.  
Using the Husimi $Q$-function we show that the phase preference is present in the long time limit only during 
non-Markovian evolution with a finite detuning. This happens due to the information backflow signifying that
non-Markovianity is a resource for quantum synchronization.  To quantify synchronization we use 
the shifted phase distribution as well as its maximal value.   From the maximal value of the shifted phase 
distribution we observe the signatures of quantum synchronization {\it viz} the Arnold tongue. In our case the 
region of synchronization is outside the tongue region and the region inside the tongue is the desynchronized
region.  This is in contrast to the results in the literature, where the synchronization is within the tongue region. 
\end{abstract}


\maketitle


\section{Introduction:}
Synchronization is the phenomenon \cite{pikovsky2001synchronization} in which a self-sustained oscillator 
adjusts its rhythm to an external weak perturbation.  It is present in a wide variety of classical systems 
like electronic circuits \cite{pecora1990synchronization}, biological neural networks 
\cite{chagnac1989synchronized,ferrari2015phase} and circadian rhythms \cite{van1928lxxii,jewett1998refinement,rompala2007dynamics}
in living systems.  All these systems need to have a stable limit cycle and also be connected to an energy source 
to sustain the oscillators indefinitely. The van der Pol oscillator is a well known system in which classical 
synchronization has been examined in detail \cite{van1926lxxxviii}.  An emerging area of research is the 
synchronization of finite dimensional quantum systems like spins and nonlinear oscillators, which has 
applications in the field of quantum computation and quantum information. Initially synchronization in the 
quantum formulation of van der Pol oscillator \cite{lee2013quantum,walter2014quantum} was studied.  Here in the 
regions far from the ground state the quantum synchronization was found to be the same as classical synchronization 
of the system under the influence of noise \cite{lee2013quantum}.  In the close to the ground state regime this 
correspondence does not exist and so we need to investigate quantum synchronization of systems with fewer 
energy levels.

Recently quantum synchronization in finite dimensional systems has been intensively studied 
\cite{koppenhofer2019optimal,parra2020synchronization,roulet2018synchronizing}.  An investigation 
on the smallest possible system that can be synchronized \cite{roulet2018synchronizing} and the conditions
under which such synchronization can happen has been carried out.  This is relevant because the quantum processor, 
fundamental to quantum computation is a collection of interconnected qubits in contact with an external environment.  
Hence in a quantum processor, synchronization of a qubit is  an  additional feature experienced due to the presence 
of other qubits, as also the driving due to an external field 
\cite{lee2013quantum,walter2014quantum,roulet2018synchronizing,xu2014synchronization}
and the environment to which the qubit is exposed \cite{karpat2019quantum,karpat2021synchronization}.  
Initially in Ref. \cite{giorgi2013spontaneous,ameri2015mutual}, qubits were suggested to be the smallest 
possible system that can be synchronized. But later works \cite{roulet2018synchronizing} claimed that 
synchronization cannot occur in dissipative two level systems. 
Subsequently it was proved that synchronization of a qubit to an external signal is possible
\cite{goychuk2006quantum,zhirov2008synchronization,eneriz2019degree}.
Generally synchronization can be classified into forced synchronization and mutual synchronization
\cite{giorgi2019transient,galve2017quantum}. In the forced synchronization or entrainment, there is a 
driving due to an external field.  The work in Ref. \cite{parra2020synchronization} discusses this feature.  

In the absence of a driving field, synchronization may emerge as a collective phenomenon due to 
coherent dynamics and this is known as mutual synchronization.  Mutual synchronization occurs due 
to the open system dynamics between qubits and an external environment.  An investigation of quantum 
synchronization induced by the environment was carried out in Ref. \cite{karpat2019quantum,karpat2021synchronization}.
In Ref. \cite{karpat2019quantum}, the authors considered two spin-$1/2$ systems in the framewor of a collision 
model.  Here they investigated the synchronization between the spins induced by the environment.  This  they  refer
to as spontaneous mutual synchronization since the synchronization between the coupled spins is influenced by their 
interaction with their respective environments.  A slightly different models of  two interacting qubits in which only one 
qubit is interacting with a disspative environment was studied in Ref.  \cite{karpat2021synchronization}.  
Here the authors explore the relation between non-Markovianity and two qubit synchronization by considering a phenomenological
Lindblad-type master equation as well as a collision model.  The results show that the information backflow and synchronization have a 
inverse relationship which can be captured by  a trade-off relation.  This raises a question as to whether a single 
quantum system exposed to a quantum environment can experience synchronization due to quantum dynamics.
In this letter we examine this question by investigating the synchronization in the time dynamics of a two level 
system in an external dissipative environment.  Our results show that there is no synchronization in the Markov
limit due to the dynamics.  In the non-Markov regime, when there is a finite detuning the two level 
systems exhibits mutual synchronization due to its interaction with the external environment.

In this letter to show phase localization we use the Husimi $Q$-function \cite{husimi1940some}.  
The amount of synchronization is measured using the shifted phase distribution $S(\phi,t)$.  
Finally to observe the characteristic feature of quantum synchronization {\it viz} the Arnold tongue, 
we use the maximum value of the shifted phase distribution.  A detailed summary of our 
results are given at the end of the letter.

\section{System and the Q-function:}
For our investigations, we consider an open quantum system composed of a two level system interacting with a dissipative 
environment.  The Hamiltonian of the qubit and the environment composite system reads: 
\begin{equation}
H = \hbar \omega_0 \sigma_{+} \sigma_{-} + \sum_k \hbar \omega_k b_k^{\dagger} b_k +
\sum_k \hbar \left( g_k \sigma_{+} b_k + g_k^{\ast} \sigma_{-} b_k^{\dagger} \right), \nonumber
\label{Spin}
\end{equation}
where $\omega_{0}$ is the transition frequency of the qubit and $\sigma_{\pm}$ are its raising and lowering operators.  The bath 
is a collection of infinite bosonic modes with creation and annihilation operators $b^{\dagger}_{k}$ and $b_{k}$.  
The coupling strength between the system and the $k^{th}$ mode of the environment with frequency $\omega_{k}$ is $g_{k}$. 
At zero temperature this model can be solved exactly.  The dynamics of the qubit is given by the time evolved reduced
density matrix: 
\begin{equation}
  \rho(t)=
  \left( {\begin{array}{cc}
  \rho_{11}(0)\ |h(t)|^{2} & \rho_{10}(0)\  h(t)\\
   \rho_{01}(0)\  h^{*}(t) & 1-\rho_{11}(0)\ |h(t)|^{2}\\
  \end{array} } \right).
\label{dynamicsmatrix}
\end{equation}
Here $h(t)$ is the time evolution function obtained from  $\dot{h}(t) = - \int_{t_{0}}^{t} d \tau  f(t-\tau) h(\tau)$
with $f(t-\tau)$ being the two time bath correlation function.  This correlation function is related to the spectral  
density $J(\omega)$ of the reservoir as $f(t-\tau) = \int d\omega J(\omega) \exp(-i (\omega_{0} - \omega) (t - \tau))$. 
The Lorentzian spectral density we consider in our work and the corresponding time evolution function $h(t)$ are: 
\begin{eqnarray}
J(\omega) &=&  \frac{1}{2 \pi} \frac{\gamma_{0} \lambda^{2}}{(\omega_{0} - \omega - \Delta)^{2} + \lambda^{2}} \nonumber \\
h(t) &=& e^{-\frac{(\lambda - i \Delta) t}{2}} \left[ \cosh \left(\frac{\Omega t}{2}\right)
       + \frac{\lambda - i \Delta}{\Omega} \sinh \left( \frac{\Omega t}{2} \right) \right], \nonumber
\end{eqnarray}
where $\lambda$ represents the spectral width of the reservoir, $\gamma_{0}$ is the coupling strength between the system 
and the bath and is related to the decay rate of the excited state of the qubit in the Markovian limit of a flat spectrum. 
The factor $\Delta = \omega_{0} - \omega_{c}$, with $\omega_{c}$ being the central frequency of the Lorentzian spectrum and
$\Omega = \sqrt{(\lambda - i \Delta)^{2}-2 \gamma_{0} \lambda}$. There are two distinct type of dynamics based on the value 
of the system-environment parameters: {\it (a)} When $\lambda > 2 \gamma_{0}$, the reservoir correlation time is very small
compared to the relaxation time of the qubit and the dynamics is Markovian. {\it (b)} For 
$\lambda < 2 \gamma_{0}$, the reservoir correlation time is larger than or of the same order as the relaxation time and we 
observe non-Markovian effects.  

\begin{figure}[h]
\includegraphics[width=\columnwidth]{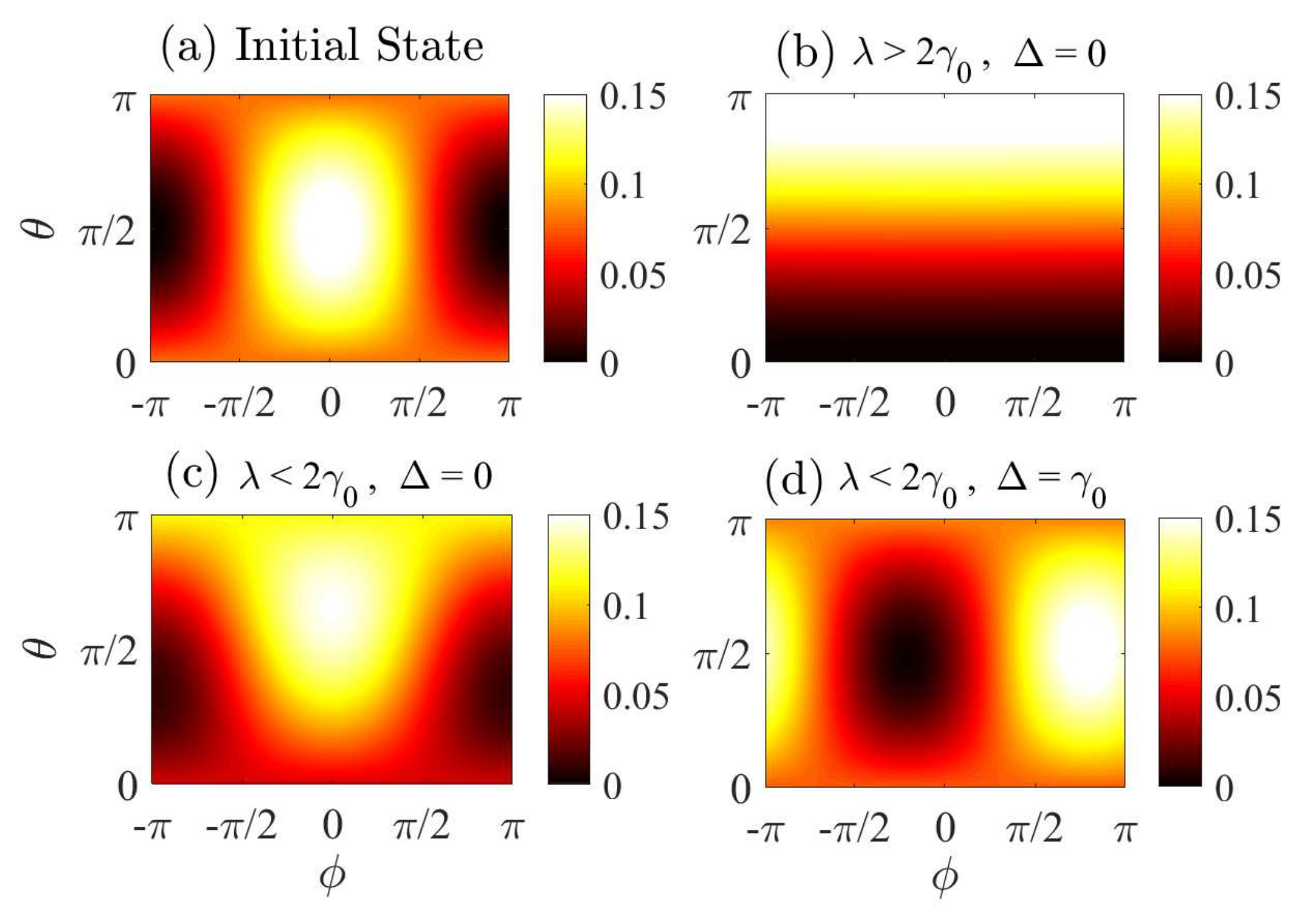}
\vskip -0.4cm
\caption{The Husimi $Q$-function $Q(\theta, \phi,t)$ of a single qubit system in the initial state 
$|+\rangle = (|0 \rangle + |1 \rangle) / \sqrt{2}$ in contact with a dissipative bath is given in the plot 
for (a) the initial time $\gamma_{0}t=0$, (b) time $\gamma_{0}t=10$ in a Markov process with $\lambda = 5 \gamma_{0}$
and $\Delta=0$, (c) time $\gamma_{0}t=10$ for non-Markov dynamics with $\lambda = 0.01 \gamma_{0}$ and $\Delta=0$,
and (d) long time limit of $\gamma_{0}t =500$, for the non-Markov process with $\Delta = \gamma_{0}$ and 
$\lambda = 0.01 \gamma_{0}$.}
\vskip -0.4cm
\label{fig1}
\end{figure}

\begin{figure}[h]
\includegraphics[width=\columnwidth]{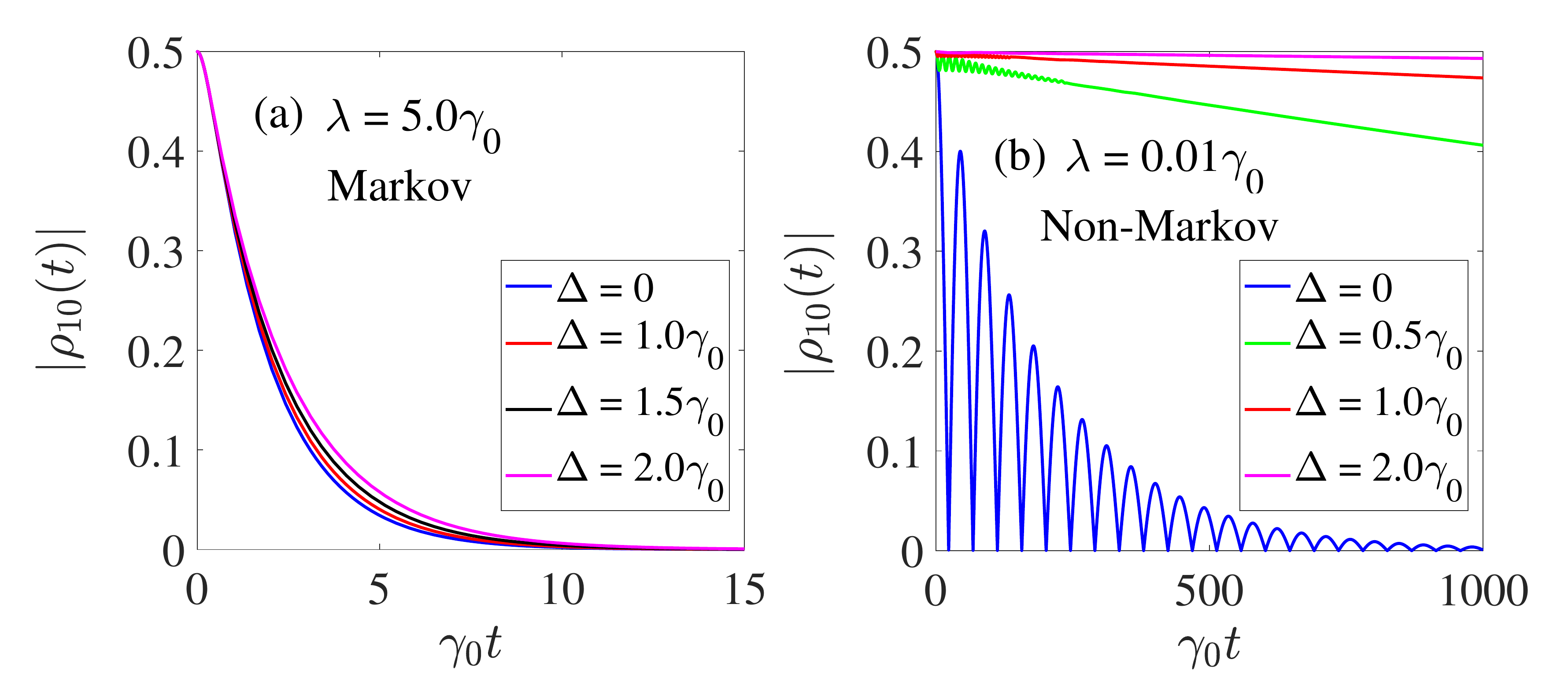} 
\vskip -0.4cm
\caption{A plot of the off-diagonal element $|\rho_{10}(t)|$ as a function of $\gamma_{0} t$ is given for 
(a) Markovian and (b) non-Markovian evolution.  }
\vskip -0.4cm
\label{fig2}
\end{figure}

To characterize the phase synchronization, we use the Husimi $Q$-function \cite{husimi1940some}, a quasi-probability 
distribution which helps to visualize the phase space of the qubit.  For our system the $Q$-function is 
\begin{eqnarray}
Q(\theta,\phi,t) = \frac{1}{2\pi} \langle \theta, \phi | \rho(t) | \theta, \phi \rangle
\label{Husimi}
\end{eqnarray}
where $|\theta, \phi \rangle =  \cos(\theta/2) |1\rangle + \sin(\theta/2) \exp (i\phi) |0\rangle$ are the 
spin-coherent states.  For the qubit system, these are the eigenstates of the spin operator 
$\sigma_{n} = \vec{n} . \hat{\sigma}$ along the axis of the unit vector $\vec{n}$, with polar co-ordinates
$\theta$ and $\phi$.  Since the spin coherent states are pure states on the Bloch sphere parametrized
by the angles $\theta$ and $\phi$, the $Q$-function gives the weights of the different pure states 
contributing to the density matrices. The  explicit form of the time dependent Husimi $Q$-function is
\begin{eqnarray}
Q(\theta,\phi,t) &=& \frac{1}{2\pi} [ \cos^2 (\theta/2) \rho_{11}(t) + \sin^2 (\theta/2) \rho_{00}(t)  \\
                 && + \sin (\theta/2) \cos (\theta/2 ) ( e^{i\phi} \rho_{10}(t) + e^{-i\phi} \rho_{01}(t)) ] \nonumber
\label{Qfunction}                 
\end{eqnarray}
The transient dynamics of the Husimi $Q$-function is shown through the plots in Figure \ref{fig1}, for the
initial state $|+\rangle = (|0 \rangle + |1 \rangle)/\sqrt{2}$.  For this single qubit state, the $Q$-distribution 
at $t=0$ is shown in Fig. \ref{fig1}(a). Here we observe the quasi-probability distribution is peaked at
$\phi=0$, indicating a non-uniform distribution of the phase of the system.  This non-uniformity implies that
the two level system has a phase preference at $\phi=0$ at $t=0$. The time evolved $Q$-function in the Markov limit 
($\lambda > 2 \gamma_{0}$) is described in the plot Fig. \ref{fig1}(b) when there is no detuning ($\Delta =0$).  Here we 
find that at $\gamma_{0} t = 10$, the $Q$-function is uniformly distributed in the $\phi$-axis and the phase preference 
is completely wiped out.  This lack of phase preference is also seen in the Markov limit of the quantum system
with a finite detuning ($\Delta \neq 0$) and the corresponding results are shown in the Appendix in Figures  7(a) and 7(b). 

Next we investigate the phase preference in the non-Markovian limit ($\lambda < 2 \gamma_{0}$), for the 
single qubit system. For $\gamma_{0} t = 10$, the time evolved $Q$-distribution of the qubit system with 
zero detuning ($\Delta =0$) is given in Fig. \ref{fig1}(c). At this time, while the phase preference is 
present in the system, it is not the same as the phase preference of the initial state.  This phase 
preference decreases with increase in the evolution time $\gamma_{0} t$ and vanishes in the long time 
limit and the plots illustrating this is given in the Appendix Fig. 8 (a).  The Husimi $Q$-function 
for the qubit system with finite detuning ($\Delta \neq 0$) in the non-Markovian limit is given in 
Fig. \ref{fig1}(d). Here we observe a dynamical phase lock in the system, i.e., only a specific region
of $\phi$ contributes to the $Q$-distribution.  This phase synchronization survives in the long time 
limit and in Fig. \ref{fig1}(d) we see the phase lock for $\gamma_{0} t = 500$, where the localized 
peak of the Husimi $Q$-function has shifted towards $\phi = \pi$.  In the Fig. 8(b) given in the appendix we 
show a series of plots displaying the evolution of the localized peak of the $Q$-distribution function 
and the system being phase locked at different $\phi$ for different times. Thus in a non-Markovian
evolution the quantum system is phase synchronized only when $\Delta \neq 0$.

To understand this we plot $|\rho_{10}(t)|$ as a function of $\gamma_{0} t$ in 
Fig. \ref{fig2}, for both Markovian and non-Markovian evolution.  For the Markovian evolution we observe from 
Fig. \ref{fig2}(a) that irrespective of the value of detuning, $|\rho_{10}(t)| \rightarrow 0$ as $t$ increases.  
Similarly for a non-Markovian dynamics when $\Delta=0$, $|\rho_{10}(t)| \rightarrow 0$ in the long-time limit. 
Thus for the Markovian evolution with any value of detuning and non-Markovian evolution with zero detuning, 
the system attains a diagonal steady state in the long-time limit.  Consequently, there is no phase localization 
due to the lack of coherence in the system \cite{koppenhofer2019optimal}.  In the case of non-Markovian evolution 
with non-zero detuning, $|\rho_{10}(t)|$ is finite even in the long time limit as shown in Fig. \ref{fig2}(b). Due to the presence of the 
off-diagonal elements (coherences) in the steady state, phase localization and consequently quantum phase 
synchronization occurs.

The phase locking of the qubit system happens due to the information backflow from the environment. 
We consider an environment which is a collection of quantum harmonic oscillators obeying bosonic 
commutation rule.  Since both the system and the bath are quantum mechanical by nature, their 
interaction leads to the synchronization of the qubit system under certain conditions.  This 
synchronization is an emergent feature due to the collective dynamics between the bath and 
the system is caused by information backflow from the bath to the system.  
The aspect of synchronization discussed in Ref. \cite{karpat2019quantum,karpat2021synchronization} 
are quite different in their origin and features. Here there are two quantum systems, out of which one system is 
influenced by an external bath which in turn synchronizes the two qubits.  The authors refer to it as 
spontaneous mutual synchronization and find that the information backflow delays the synchronization. 
These two types of synchronizations are quite different in their origin and features \cite{galve2017quantum}.

\begin{figure}[h]
\includegraphics[width=\columnwidth]{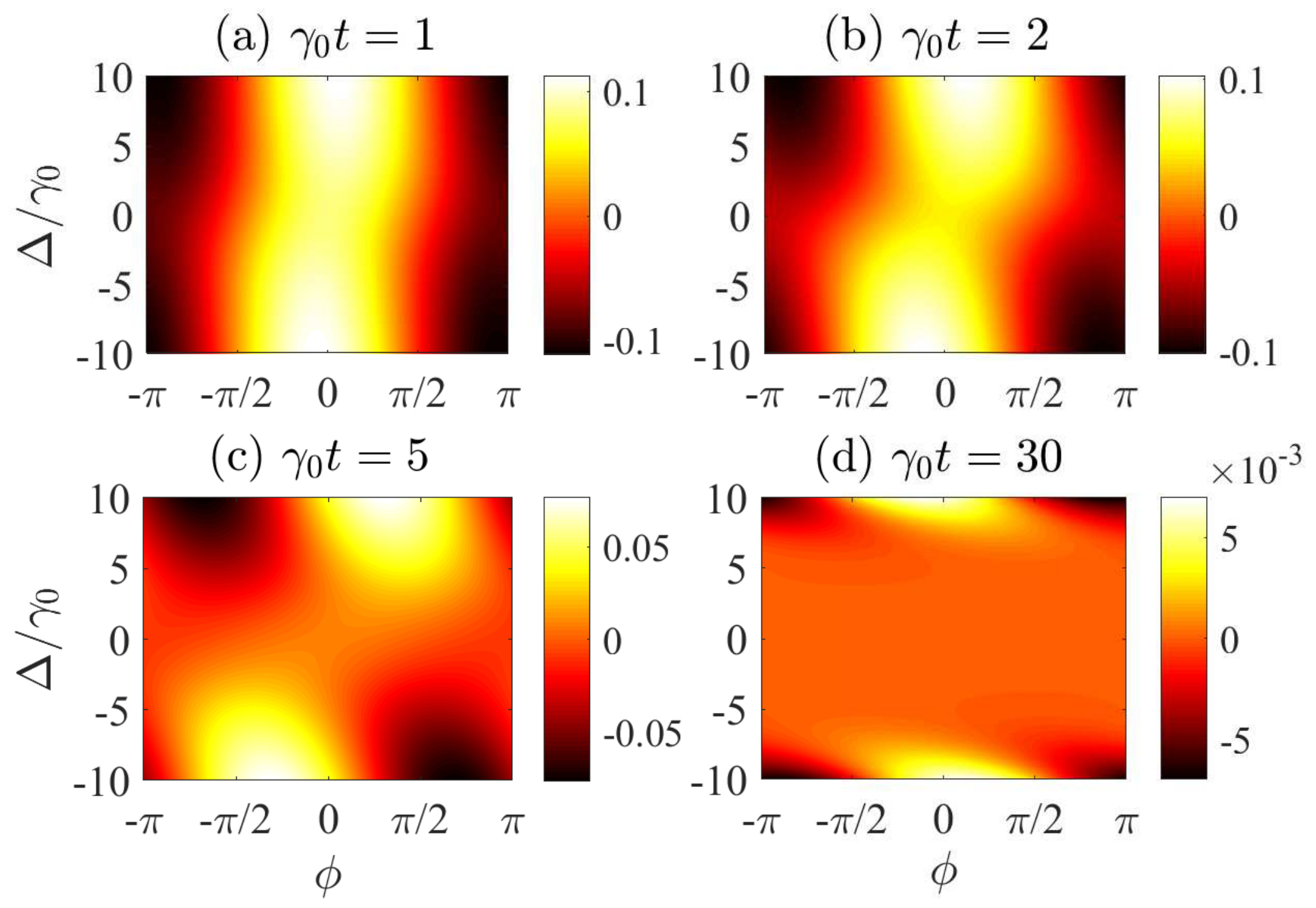} 
\vskip -0.4cm
\caption{For the Markovian dynamics we show a contour plot describing the shifted phase distribution
$S(\phi,t)$ as a function of $\Delta$ and $\phi$ for (a) $\gamma_{0} t = 1$, (b) $\gamma_{0} t = 2$,  
(c) $\gamma_{0} t = 5$ and (d) $\gamma_{0} t = 30$.  The spectral width $\lambda =5 \gamma_{0}$ for 
all the plots.}
\vskip -0.4cm
\label{fig3}
\end{figure}

\section{ Synchronization measure and Arnold's tongue:}
The Husimi Q-function explicitly shows the phase preference in the system. By integrating over the 
angular variable `$\theta$' we can determine the phase distribution $P(\phi, \rho)$ for a given state 
$\rho$.  For a limit cycle state $\rho_{0}$, $P(\phi, \rho_{0}) = 1/2 \pi$ indicating a uniform phase distribution.  
Hence the synchronization of the qubit system can be measured using the shifted phase distribution
\begin{equation}
 S(\phi,t) = \int_{0}^{\pi} d \theta \sin \theta  Q(\theta, \phi,t) - \frac{1}{2 \pi}.
\end{equation}
This function is zero if and only if there is no phase preference in the system implying the absence of 
phase synchronization.  Evaluating the integral over the angular variable $\theta$, we find 
\begin{equation}
    S(\phi,t) = \frac{1}{8} \left[ \rho_{10}(t) \exp( i \phi) + \rho_{01}(t) \exp(-i \phi) \right].
\end{equation}
To estimate quantum synchronization, we plot $S(\phi,t)$ as a function of $\Delta$ and $\phi$ for different
values of time. The plots for the Markovian dynamics ($\lambda > 2 \gamma_{0}$) for the time values 
$\gamma_{0} t = 1$, $\gamma_{0} t = 2$, $\gamma_{0} t = 5$ and $\gamma_{0} t = 30$ are shown through the 
plots in Fig. \ref{fig3} (a), \ref{fig3} (b), \ref{fig3} (c), and \ref{fig3} (d) respectively. 
From the series of plots in Fig. \ref{fig3} (b) - \ref{fig3} (d) we observe the formation of small region 
around $\Delta=0$ with no phase synchronization.  The size of this region is proportional to $\gamma_{0} t$
and in the long time limit, there is no phase synchronization for physically acceptable values of detuning
as can be seen in Fig. \ref{fig3}(d).

\begin{figure}[h]
\includegraphics[width=\columnwidth]{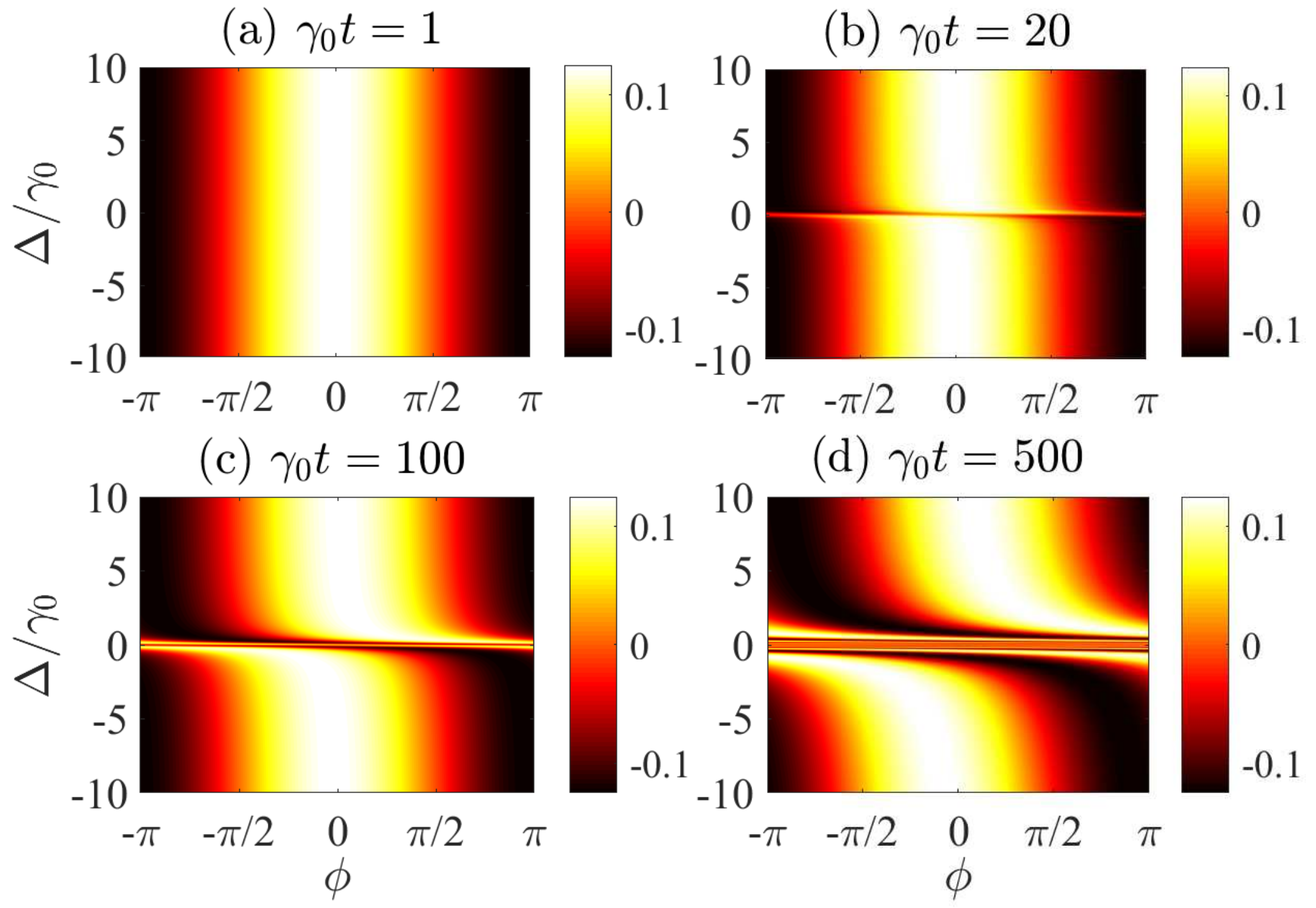}
\vskip -0.35cm
\caption{ The contour plot of the shifted phase distribution $S(\phi,t)$ as a function of $\Delta$ and 
$\phi$ for the non-Markovian dynamics for $\lambda =0.01  \gamma_{0}$  is given for 
(a) $\gamma_{0} t = 1$, (b) $\gamma_{0} t = 20$,  (c) $\gamma_{0} t = 100$ and (d) $\gamma_{0} t = 500$.}
\vskip -0.35cm
\label{fig4}
\end{figure}

The non-Markovian dynamics of the phase synchronization is shown in Fig. \ref{fig4}(a)-(d) using the shifted 
phase distribution.  The different evolution times considered in Fig. \ref{fig4}(a), 
\ref{fig4}(b), \ref{fig4}(c), and \ref{fig4}(d) are $\gamma_{0} t =1, 20, 100$ and $500$ respectively.  
At $\gamma_{0} t =1$ (\ref{fig4} (a)), the shifted phase distribution is peaked around $\phi=0$,
implying a synchronization in that region. In the neighborhood of $\phi = \pm \pi/2$, the function 
$S(\phi,t) = 0$ signifying a lack of synchronization.  Beyond this region and close to 
$\phi = \pm \pi$ the shifted phase distribution $S(\phi,t) = - 1/8$ due to antisynchronization. This 
behavior is uniformly seen for all values of detuning. As the system evolves, we observe the 
emergence of a region with no synchronization ($S(\phi,t) = 0$) around $\Delta = 0$ and this 
region transitions from being a line of negligible width in Fig. \ref{fig4} (b) to a narrow band
of finite width in the detuning parameter as seen in Fig. \ref{fig4} (d).  So we conclude that in a 
non-Markovian evolution phase synchronization happens only for finite value of detuning and for zero 
or negligible detuning there is no phase synchronization.

\begin{figure}[h]
\includegraphics[width=\columnwidth]{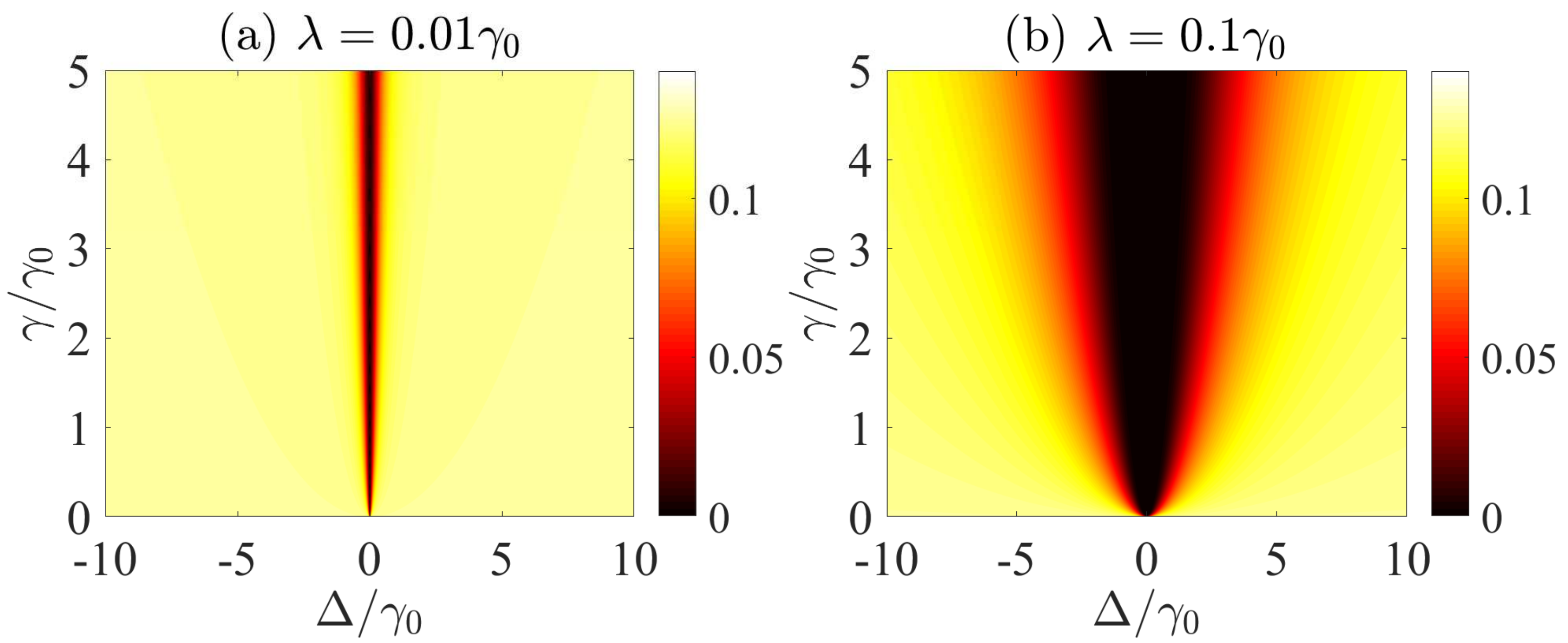} 
\caption{This plot describes the maximal value of the shifted phase distribution $S_{m}(t)$ as a function of 
the detuning parameter $\Delta$ and coupling strength $\gamma_{0}$ for the spectral width 
(a) $\lambda = 0.01 \gamma_{0}$ and (b) $\lambda = 0.1 \gamma_{0}$. We consider $\gamma_{0} t = 500$
for both the plots.}
\vskip -0.3cm
\label{fig5}
\end{figure}

The maximum of the shifted phase distribution $S_{m}(t) = \max  S(\phi,t)$ can also be used to 
characterize phase synchronization. We plot $S_{m}(t)$ in Fig. \ref{fig5}, as a function of the detuning 
parameter ($\Delta$) and the system-bath coupling strength ($\gamma$). To vary the coupling strength, we 
consider the spectral density $J(\omega)$ with coupling parameter $\gamma$ which varies in units of $\gamma_{0}$.
In Fig. \ref{fig5} (a) we consider a spectral width $\lambda = 0.01 \gamma_{0}$ and in Fig. \ref{fig5} (b) we consider 
$\lambda=0.1 \gamma_{0}$ and in both instances we consider $\gamma_{0} t =500$.  We observe the formation of a Arnold tongue 
in both the plots.  But in contrast to the results observed in the literature 
\cite{koppenhofer2019optimal,parra2020synchronization,roulet2018synchronizing,roulet2018quantum}
we find there is no synchronization in the tongue region, rather it is present only in the 
region outside of the tongue like formation. The central line of the tongue region is at $\Delta=0$ and the 
width of the tongue increases with the spectral width ($\lambda$). This implies that the region of synchronization increases
with the non-Markovian behavior. So in our model, the Arnold tongue marks out the unsynchronized region.   
\begin{figure}[h]
\includegraphics[width=\columnwidth]{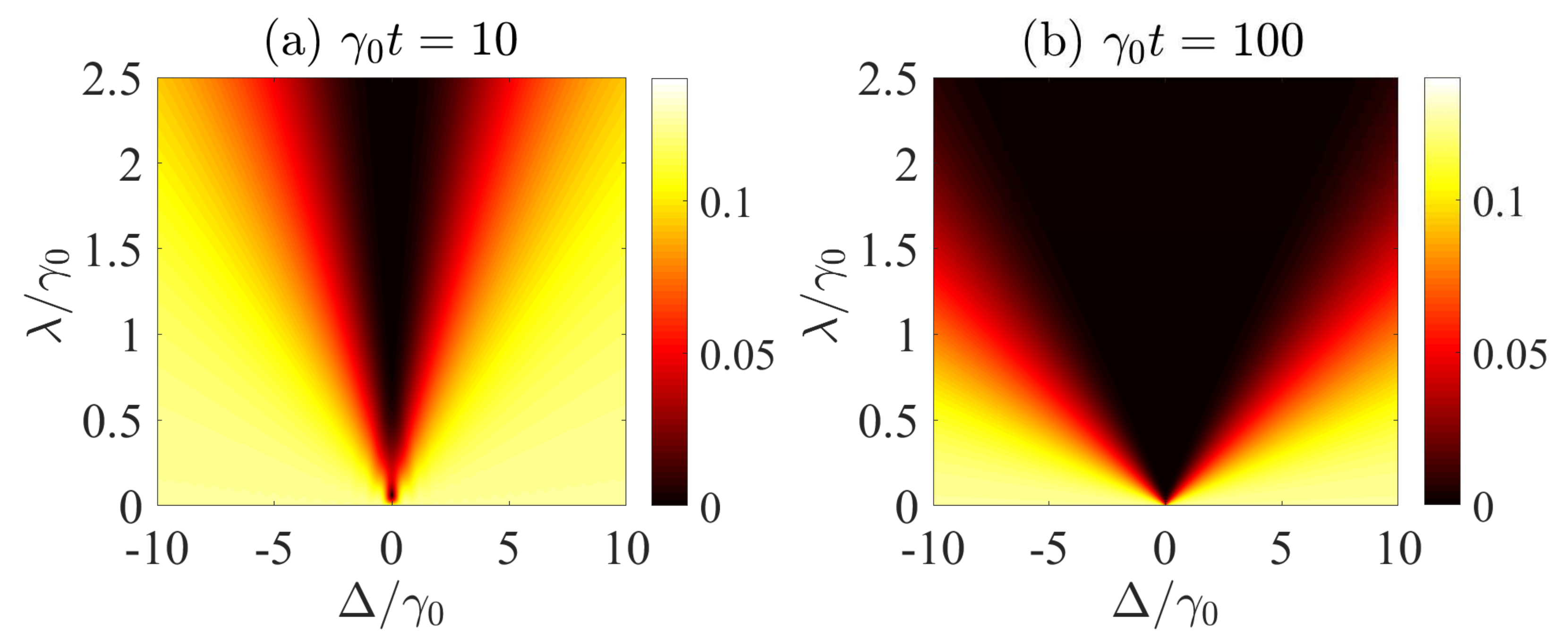} 
\caption{ A contour plot of the maximal shifted phase distribution $S_{m}(t)$ Vs detuning parameter $\Delta$ 
Vs Spectral width $\lambda$ is given for (a) $\gamma_{0} t = 10$ and (b) $\gamma_{0} t = 100$. }
\label{fig6}
\end{figure}
The quantum synchronization in the system is dependent on the parameters {\it viz} detuning ($\Delta$), 
the system-environment coupling strength ($\gamma$) and the spectral width ($\lambda$).  In Fig. \ref{fig5},
we investigated the variation of $S_{m}(t)$ as a function of $\Delta$ and $\gamma$. To have a complete picture
about the synchronization we also plot $S_{m}(t)$ as a function of $\Delta$ and $\lambda$ in Fig. \ref{fig6} for 
different values of $\gamma_{0} t$. Here too we observe a triangular unsynchronized region similar to an 
Arnold tongue. and outside this region the system is synchronized. From the plots Fig. \ref{fig6} (a) and 
\ref{fig6} (b) we can see that the unsynchronized region increases with time.  From Fig. \ref{fig6} (b) we 
realize that to have synchronization in the long time limit, we need to choose the spectral width $\lambda$ 
to be small.

\section{Conclusion:}
In  this letter we show that quantum mutual synchronization can be induced by  information backflow.  For 
this we consider a two level system exposed to an external environment modelled by an infinite bosonic modes. 
We consider a Lorentzian spectrum for the bath and examine the system for quantum synchronization under both 
Markovian and non-Markovian dynamics.  We use the Husimi $Q$-function to plot the phase space of the qubit 
and we notice that when the system is under Markovian dynamics the initial phase preference is lost and consequently 
there is no transient phase synchronization.   Under non-Markovian evolution for finite detuning, there is a phase 
preference even in the long-time limit, signalling the presence of a phase synchronization whose occurrence is due 
to the coherent dynamics and the information backflow into the system.  This synchronization is an emergent phenomenon 
due to the coherent dynamics.  In earlier works 
\cite{karpat2019quantum,karpat2021synchronization} it was established that information backflow prevents or delays 
the onset of synchronization between quantum systems.  Here, the main focal point of these investigations 
\cite{karpat2019quantum,karpat2021synchronization} was on how the environment influences the synchronization 
between two quantum systems.  In our work, we show that the environment is capable of directly inducing phase 
synchronization in a qubit.  Thus we demonstrate that the information backflow has a positive effect on the collective 
synchronization and this contradicts the role of information backflow on spontaneous mutual synchronization in a 
two qubit system as described in Ref. \cite{karpat2019quantum,karpat2021synchronization}.

To quantify synchronization we use the shifted phase distribution and plot it as a function of the 
detuning and phase.  In the Markov case there is no synchronization for physically acceptable values 
of detuning.  During the non-Markovian evolution we observe synchronized regions in the long-time 
limit for finite values of detuning.  Finally. we plot the maximum of the shifted phase distribution 
in two different ways: (a) By varying the detuning and the interaction strength and (b) through a 
variation of detuning and the spectral density width.  In both the cases we observe the formation of 
Arnold tongue but in our study the qubit system is synchronized in the region outside the tongue formation
and desynchronized within the tongue region.  This is in contrast to the studies of synchronization so far 
\cite{koppenhofer2019optimal,parra2020synchronization,roulet2018synchronizing,roulet2018quantum} 
where the synchronization happens within the tongue region.

\section{Acknowledgements:}
Md. Manirul Ali was supported by the Centre for Quantum Science and Technology, Chennai Institute of 
Technology, India, vide funding number CIT/CQST/2021/RD-007. Po-Wen Chen was supported by the Division of 
Physics, Institute of Nuclear Energy Research, Taiwan. Chandrashekar Radhakrishnan was supported in part 
by a seed grant from IIT Madras to the Centre for Quantum Information, Communication and Computing.


\appendix

\section{The model}

\begin{figure*}[t]
\centering{\rotatebox{0}{\resizebox{17.5cm}{6.5cm}{\includegraphics{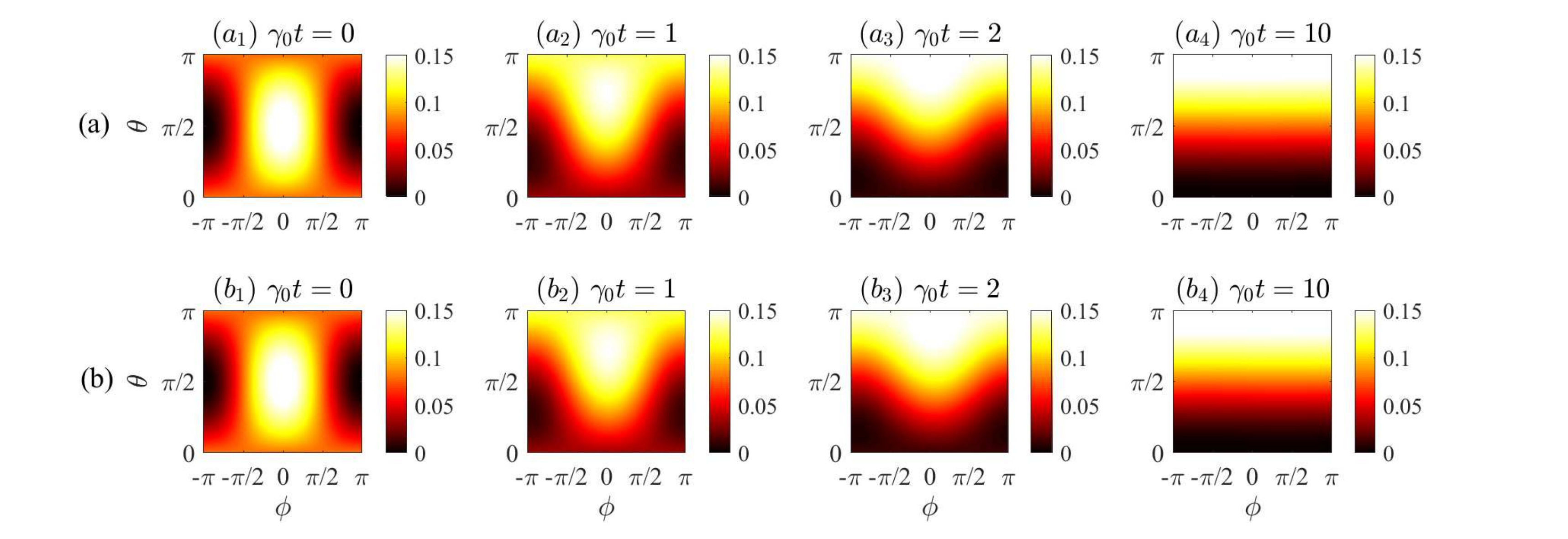}}}}
\caption{\label{sfig1} Temporal evolution of the Husimi distribution function $Q(\theta,\phi,t)$
is shown in the Markov regime ($\lambda > 2 \gamma_0$) for (a) $\Delta=0$ and (b) $\Delta=\gamma_0$.
We have taken the spectral width $\lambda=5\gamma_0$.}
\end{figure*}

For our investigations we consider a dissipative two-level quantum system spontaneously decaying under 
the influence of an environment with a continuum of modes. The total Hamiltonian of the system plus environment is given by
\begin{eqnarray}
\label{TotH}
H = H_{S} + H_{E} + H_{I},
\end{eqnarray}
where the system Hamiltonian $H_S=\hbar \omega_0 \sigma_{+}\sigma_{-}$ is expressed in terms of the
spin raising and lowering operators $\sigma_{+} = |1\rangle \langle 0|$ and $\sigma_{-} = |0 \rangle \langle 1|$. 
The qubit has a transition frequency $\omega_0$ between the states $|0\rangle$ and $|1\rangle$. 
The environment Hamiltonian $H_E = \sum_k \hbar \omega_k b_k^{\dagger} b_k$
describes a collection of Bosonic modes, where $b_k^{\dagger}$ and $b_k$ are the corresponding creation and 
annihilation operators of the kth mode with frequency $\omega_k$. The interaction between the system
and the Bosonic bath is described by the Hamiltonian
\begin{eqnarray}
\label{IntH}
H_{I} = \sum_k \left( g_k \sigma_{+} \otimes b_k + g_k^{\ast} \sigma_{-} \otimes b_k^{\dagger} \right).
\end{eqnarray}
We go to the interaction picture with respect to $H_0 = H_S + H_E$, then the Schr\"{o}dinger equation for the 
total system-plus-environment state is given by
\begin{eqnarray}
\label{Schrod}
\frac{d}{dt} | \Psi (t)\rangle = -\frac{i}{\hbar} H_{I}(t)  | \Psi (t) \rangle,
\end{eqnarray}
where
\begin{eqnarray}
\label{IntH2}
H_I (t) &=& e^{\frac{i}{\hbar} H_0 t } ~H_{I}~ e^{-\frac{i}{\hbar}H_0 t}, \\
\nonumber
&=& \hbar \left[ \sigma_{+} \otimes B(t) +  \sigma_{-} \otimes B^{\dagger}(t) \right].
\end{eqnarray}
with $B(t) = \sum_k g_k  e^{i(\omega_0 - \omega_k) t}~b_k$. We start with an initial state 
$|\Psi(0)\rangle = \left( d_0 |0\rangle + d_1(0) |1\rangle \right) \otimes |0\rangle_{E}$, the 
environment is initially in the vacuum state $|0\rangle_E=|000....0....0\rangle$.
The exact time evolution of $|\Psi(0)\rangle$ is given by \cite{scully_zubairy_1997}
\begin{align}
\nonumber
|\Psi(t)\rangle =  & ~d_0 |0\rangle \otimes |0\rangle_{E} + d_1(t) |1\rangle \otimes |0\rangle_{E}\\
& + \sum_k d_k (t) |0\rangle \otimes |k\rangle_{E},
\label{tPsi}
\end{align}
where $|k\rangle_{E}=b_k^{\dagger}|0\rangle_{E}=|000....1_{k} ....0\rangle$ is the state with one photon
only in the mode $k$. The time evolved state is confined to the subspace spanned by the vectors
$|0\rangle \otimes |0\rangle_{E}$, $|1\rangle \otimes |0\rangle_{E}$, and $|0\rangle \otimes |k\rangle_{E}$.
This is because the Schr\"{o}dinger equation is generated by the Hamiltonian $H_{I}(t)$ that conserves the 
total particle number. The amplitudes $d_1(t)$ and $d_k(t)$ depend on time, while $d_0$ is constant in time
because $H_{I}(t)  |0\rangle \otimes |0\rangle_{E}=0$. Substituting $|\Psi(t)\rangle$ from Eq.~(\ref{tPsi})
into the Schr\"{o}dinger equation (\ref{Schrod}), one can obtain
\begin{subequations}
\begin{align}
\label{d1}
&\frac{d}{dt} d_1(t) = - i \sum_k g_k e^{i(\omega_0-\omega_k)t} d_k(t), \\
\label{dk}
&\frac{d}{dt} d_k(t) = - i g^{\ast}_k e^{-i(\omega_0-\omega_k)t} d_1(t).
\end{align}
\end{subequations}
Integrating Eq.~(\ref{dk}) with the initial condition $d_k(t)=0$ at $t=0$, we have
\begin{eqnarray}
\label{dkt}
d_k(t) = - i g^{\ast}_k \int_{0}^{t} e^{-i(\omega_0-\omega_k)\tau} d_1(\tau) d\tau.
\end{eqnarray}
We substitute this solution for $d_k(t)$ in Eq.~(\ref{d1})
\begin{eqnarray}
\label{d1t}
\frac{d}{dt} d_1(t) = - \int_{0}^{t} \sum_k |g_k|^2   e^{i(\omega_0-\omega_k)(t-\tau)} d_1(\tau) d\tau,
\end{eqnarray}
which is expressed as an integrodifferential equation for $d_1(t)$ and the equation reads: 
\begin{eqnarray}
\label{amp1}
\frac{d}{dt} d_1(t) = - \int_{0}^{t} d\tau f(t-\tau) d_1(\tau),
\end{eqnarray}
where $f(t-\tau)$ $=\sum_k |g_k|^2 e^{i(\omega_0-\omega_k)(t-\tau)}$ is the two-time correlation function
$\langle 0| B(t) B^{\dagger}(\tau) |0\rangle_{E}$ of the reservoir. If the reservoir spectrum is continuous, the 
two-time correlation function can be expressed through the spectral density $J(\omega)$ of the environment 
as $f(t-\tau) =  \int d\omega J(\omega) e^{i(\omega_0 - \omega)(t-\tau)}$. The correlation function $f(t-\tau)$
characterize all the non-Markovian back-action effects between the system and the reservoir. Here $J(\omega) = 
\sum_k |g_k|^2 \delta(\omega-\omega_k) = \mathcal{P}(\omega) |g(\omega)|^2$  and $\mathcal{P}(\omega)$ is 
the density of states of the reservoir and $g(\omega)$ is the frequency-dependent coupling  between the system 
and the bath. The spectral density $J(\omega)$ contain the information about the frequency distribution of environmental modes and 
also the coupling between the system and the environment. Then the reduced density operator of the
system is $\rho (t) = {\rm Tr}_E \{ | \Psi (t) \rangle \langle \Psi (t)| \}$ is determined by the function $d_1(t)$ 
and is 
\begin{align}
\rho(t) = & |d_1(t)|^2 ~ |1\rangle \langle 1| + d_1(t) d^{\ast}_{0} ~ |1\rangle \langle 0| \\
\nonumber
& d_{1}^{\ast}(t) d_0 ~ |0\rangle \langle 1|  +  \left(1- |d_1(t)|^2 \right) |0\rangle \langle 0|.
\end{align}
One can define the time evolution function $h(t)$ as the solution of the equation
\begin{eqnarray}
\label{ht}
\frac{d}{dt} h(t) = - \int_{0}^{t} d\tau f(t-\tau) h(\tau),
\end{eqnarray}
with the initial value $h(0)=1$. Then the amplitude $d_1(t) = h(t) d_1(0)$ and the time dependent density
matrix elements for the qubit are given by
\begin{eqnarray}
\label{rho}
\rho_{11}(t) &=& \langle 1| \rho(t) |1 \rangle = |d_1(t)|^2 = |h(t)|^2 |d_1(0)|^2, \\
\nonumber
\rho_{10}(t) &=& \langle 1| \rho(t) |0 \rangle = d_1(t) d^{\ast}_{0} = h(t) d_1(0) d^{\ast}_{0}, \\
\nonumber
\rho_{01}(t) &=& \langle 0| \rho(t) |1 \rangle = d_{1}^{\ast}(t) d_0 = h^{\ast}(t) d^{\ast}_{1}(0) d_0, \\
\nonumber
\rho_{00}(t) &=& \langle 0| \rho(t) |0 \rangle = \left(1- |d_1(t)|^2 \right) = 1 -  |h(t)|^2 |d_1(0)|^2.
\end{eqnarray}
Hence the dissipative dynamics of the qubit is described by the time evolved reduced density matrix
\begin{equation}
\rho(t)=
\left( {\begin{array}{cc}
\rho_{11}(0)\ |h(t)|^{2} & \rho_{10}(0)\  h(t)\\
\rho_{01}(0)\  h^{*}(t) & 1-\rho_{11}(0)\ |h(t)|^{2}\\
\end{array} } \right),
\label{denmat}
\end{equation}
where $\rho_{11}(0)$ $=|d_1(0)|^2$, $\rho_{10}(0)$ $=d_1(0) d^{\ast}_{0}$, and 
$\rho_{01}(0)$ $=d^{\ast}_{1}(0) d_0$. The above model of dissipation for a two-level quantum system
is used in the context of non-Markovian dynamics 
\cite{bellomo2007non,tong2010decoherence,li2010entanglement,breuer2016colloquium} of open quantum systems. 
This exactly soluble model is also applied to investigate the measure of non-Markovianity 
\cite{breuer2009measure,addis2014comparative,hou2015non,pineda2016measuring}
for two-level open quantum systems.

\begin{figure*}[t]
\centering{\rotatebox{0}{\resizebox{17.5cm}{6.0cm}{\includegraphics{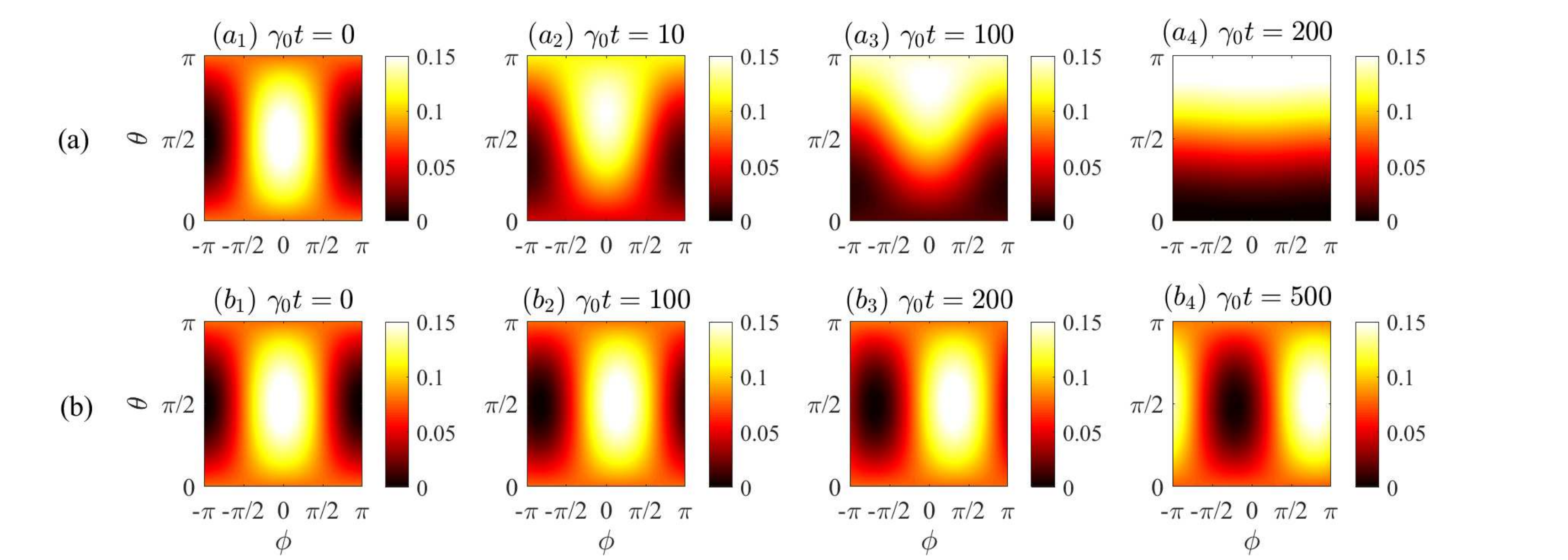}}}}
\caption{\label{sfig2} Temporal evolution of the Husimi distribution function $Q(\theta,\phi,t)$
is shown in the non-Markov regime ($\lambda < 2 \gamma_0$) for (a) $\Delta=0$ and (b) $\Delta=\gamma_0$.
We have taken the spectral width $\lambda=0.01\gamma_0$.}
\end{figure*}

\section{Husimi distribution}

To visualize and characterize the phase synchronization behavior of a two-level quantum system, we use the
Husimi $Q$ distribution  \cite{husimi1940some,roulet2018quantum} defined by
\begin{eqnarray}
Q(\theta,\phi,t) = \frac{1}{2\pi} \langle \theta, \phi | \rho(t) | \theta, \phi \rangle,
\label{Husimi}
\end{eqnarray}
where the spin-coherent states $|\theta, \phi \rangle$ are the eigenstates of the spin operator $\sigma_n = \vec{n}.\hat{\sigma}$
along the unit vector $\vec{n}$ defined by the polar coordinates $\theta$ and $\phi$. These are the pure states
$| \theta, \phi \rangle = \cos(\theta/2) |1\rangle + \sin(\theta/2) \exp (i\phi) |0\rangle$, representing a point on the Bloch sphere.
Once the time evolved density matrix elements are determined, it is easy to obtain the time dynamics of $Q$ distribution as a
function of $\theta$, $\phi$ and $t$ as follows: 
\begin{widetext}
\begin{eqnarray}
\nonumber
Q(\theta,\phi,t) &=& \frac{1}{2\pi} {\rm Tr} \left( | \theta, \phi \rangle \langle \theta, \phi | \rho(t) \right) =
\frac{1}{2\pi} {\rm Tr} \left[
\left(\begin{array}{cc}
\cos^2 (\theta/2)  & \sin (\theta/2) \cos (\theta/2) e^{-i\phi} \\
\sin (\theta/2) \cos (\theta/2) e^{i\phi} & \sin^2 (\theta/2) \\
\end{array}
\right)
\left(\begin{array}{cc}
\rho_{11}(t)  & \rho_{10}(t) \\
\rho_{01}(t) & \rho_{00}(t) \\
\end{array}
\right) \right], \\
&=& \frac{1}{2\pi} \left[ \cos^2 (\theta/2) \rho_{11}(t) +  \sin (\theta/2) \cos (\theta/2) e^{-i\phi} \rho_{01}(t)
+ \sin (\theta/2) \cos (\theta/2) e^{i\phi} \rho_{10}(t) + \sin^2 (\theta/2) \rho_{00}(t) \right].
\label{Qfun_def}
\end{eqnarray}
\end{widetext}

We start with an initial spin-coherent state corresponding to $\theta=\pi/2$ and $\phi=0$. Figs.~\ref{sfig1} and  \ref{sfig2} show
the time-dependent transient dynamics of $Q(\theta,\phi,t)$ for Markov and non-Markov evolution respectively.
In Fig.~\ref{sfig1}, the Husimi $Q$ function is shown
in the Markov regime ($\lambda > 2 \gamma_0$) for two different values of the detuning (a) $\Delta=0$ and
(b) $\Delta=\gamma_0$ respectively. For zero detuning (Fig.~\ref{sfig1}) we see that initially at time $t=0$,
$Q$ function is nonuniform in phase and the distribution is peaked at $\phi=0$. As time passes this phase preference
of the Husimi distribution diminishes and is eventually wiped out at a time $\gamma_0 t=10$, after which the $Q$
function becomes uniformly distributed along $\phi$-axis. Next in Fig.~\ref{sfig1}, we consider a nonzero
finite value of the detuning $\Delta=\gamma_0$. Even for this nonzero value of detuning (Fig.~\ref{sfig1}), the transient
dynamics of $Q$ distribution is quite similar to that of zero detuning case as long as the reservoir parameters are set in the
Markov regime. The phase information of the state smears out quickly after a short interval of the system-environment
interaction.

The non-Markovian counterpart of the dynamics is shown in Fig.~\ref{sfig2} for the reservoir parameter with $\lambda < 2 \gamma_0$.
For zero detuning ($\Delta=0$) we still observe a loss of phase information as time progresses, although this loss of phase preference in $Q$
function is delayed or slowed down (Fig.~\ref{sfig2} (a)) here in the non-Markov regime. 
Finally, the nonuniform phase preference is ultimately lost and the $Q$ distribution gets delocalized with respect to
$\phi$ at $\gamma_0 t=200$. The situation is dramatically changed in the nonzero detuning ($\Delta \ne 0$). In Fig.~\ref{sfig2} (b),
we show the non-Markov dynamics of the $Q$ function for a nonzero finite value of the detuning.
In the non-Markov regime ($\lambda < 2 \gamma_0$) with nonzero detuning ($\Delta=\gamma_0$), we see a dynamical
phase-locking for the two level system. The qubit is phase locked in the sense that the $Q$ distribution is mostly contributed from
a specific $\phi$ region. The localized peak of the Husimi $Q$ distribution gradually shifts from $\phi=0$ towards $\phi=\pi$ as
time progress. Hence in the non-Markov regime for this finite detuning case ($\Delta=\gamma_0$), we observe here a transient
phase-synchronization characterized by a single peak $Q$-distribution.

\section{synchronization measure and Arnold's tongue}

We have identified phase synchronization for a two-level quantum system through a phase-space quasiprobability 
distribution, called Husimi $Q$-function. Husimi $Q$-function precisely demonstrates the synchronization in the 
qubit system. Moreover, we adopt \cite{koppenhofer2019optimal,parra2020synchronization,roulet2018synchronizing} 
a measure of synchronization to characterize the strength of the phase preference. Integrating the $Q$-function over the angular 
variable `$\theta$' one can obtain a phase distribution $P(\phi, \rho)$ for a given state $\rho$. Corresponding to a limit cycle state, 
the phase distribution $P(\phi, \rho) = 1/2 \pi$ that indicates a uniform phase distribution. Then the synchronization of the two-level 
quantum system can be measured using the shifted phase distribution
\begin{equation}
S(\phi,t) = \int_{0}^{\pi} d \theta \sin \theta ~Q(\theta, \phi,t) - \frac{1}{2 \pi}.
\label{SFun1} 
\end{equation}
This function is zero in the absence of synchronization when there is no phase preference in the system. 
A positive maximum of this measure indicates a complete phase locking, while a negative maximum 
implies the anti-synchronization. Substituting $Q(\theta, \phi,t)$ from Eq.~(\ref{Qfun_def}) in Eq.~(\ref{SFun1}),
we evaluate the integral over the angular variable $\theta$
\begin{align}
S(\phi,t) &= \frac{1}{2\pi} \Big[ ~\rho_{11}(t) \int_0^{\pi} \cos^2 (\theta/2) \sin \theta d\theta - 1 \\
\nonumber
& \hskip 0.8cm + \rho_{00}(t) \int_0^{\pi} \sin^2 (\theta/2) \sin \theta d\theta \\
\nonumber
& \hskip 0.8cm + \rho_{10}(t) ~e^{i\phi} \int_0^{\pi} \sin (\theta/2) \cos (\theta/2) \sin \theta \\
\nonumber
& \hskip 0.8cm + \rho_{01}(t) ~e^{-i\phi}\! \int_0^{\pi}\! \sin (\theta/2) \cos (\theta/2) \sin \theta \Big].
\label{SFun2}
\end{align}
By performing the above integral and using the trace preserving property $\rho_{11}(t)+\rho_{00}(t)=1$, we have
\begin{equation}
S(\phi,t) = \frac{1}{8} \Big[ \rho_{10}(t) e^{i\phi} + \rho_{01}(t) e^{-i\phi} \Big].
\label{SFun3}
\end{equation}
To estimate quantum synchronization, we plot $S(\phi,t)$ as a function of $\Delta$ and $\phi$ for different
values of time (see Figure 3 and Figure 4 of the main manuscript). The plots for the Markovian ($\lambda > 2 \gamma_{0}$) 
dynamics of $S(\phi,t)$ are shown in Figure 3 of the main manuscript, where we observe the formation of small region
around $\Delta=0$ with no phase synchronization. The size of this region is proportional to $\gamma_{0} t$
and in the long time limit, there is no phase synchronization for physically acceptable values of detuning. 
The non-Markovian dynamics (for $\lambda < 2 \gamma_0$) of the phase synchronization is also shown 
(see Figure 4 of the manuscript) using this shifted phase distribution $S(\phi,t)$. We see that quantum phase 
synchronization region is enhanced in the non-Markovian regime, the function achieves a maximum value ($S(\phi,t) = 1/8$)
if one avoids a narrow band region around ($\Delta = 0$) just by introducing a small amount of detuning. The narrow
band region in $S(\phi,t)$ around $\Delta = 0$ signifies a no-synchronization ($S(\phi,t) = 0$) zone. We also observed 
anti-synchronization ($S(\phi,t) = - 1/8$) region for some specific range of values of $\phi$. So we conclude that 
for the Markov evolution there is no synchronization in the long-time limit for any value of detuning. For non-Markovian 
evolution phase synchronization happens only for finite value of detuning and for zero or negligible detuning there is no 
phase synchronization.

The maximum of the shifted phase distribution $S_{m}(t) = \max  S(\phi,t)$ also provides an
alternative measure of phase synchronization. The maximum value of $S(\phi,t)$ can be calculated
as follows. From Eq.~(\ref{SFun3}), we have
\begin{equation}
S(\phi,t) = \frac{1}{4} ~{\rm Re} \Big[ \rho_{10}(t) e^{i\phi} \Big].
\label{SFun4}
\end{equation}
The density matrix element $\rho_{10}(t)$ is in general a complex number that can be expressed as 
$\rho_{10}(t)=r(t) e^{i\theta(t)}$, where $r(t)=\sqrt{x^2(t) + y^2(t)}$ and $\theta(t)=\tan^{-1}(y(t)/x(t))$. Here
$x(t)={\rm Re}~[\rho_{10}(t)]$ and $y(t)={\rm Im}~[\rho_{10}(t)]$. Then from Eq.~(\ref{SFun4}), we have
\begin{eqnarray}
\hskip -0.5cm S(\phi,t) = \frac{1}{4} ~{\rm Re} \Big[ r(t) e^{i\theta(t)} e^{i\phi} \Big] = \frac{1}{4} r(t) \cos \Big(\theta(t) + \phi \Big).
\label{SFun5}
\end{eqnarray}

To find the maximum value of $S(\phi,t)$ with respect to the variable $\phi$
\begin{align}
\frac{\partial S}{\partial \phi} = -\frac{1}{4} r(t) \sin \Big(\theta(t) + \phi \Big) = 0.
\label{SDiff}
\end{align}
Hence the function $S(\phi,t)$ will have a maximum or minimum value at $\theta(t)+\phi=n\pi$, where
$n$ is an integer. One can easily check that the function $S(\phi,t)$ has a maximum at $\phi=-\theta(t)$
for which the second derivative 
\begin{eqnarray}
\nonumber
\frac{\partial^2 S}{\partial \phi^2}\Big|_{\phi=-\theta(t)} &=& -\frac{1}{4} r(t) \cos \Big(\theta(t) + \phi \Big) \Big|_{\phi=-\theta(t)}, \\
&=& -\frac{r(t)}{4},
\label{S2Diff}
\end{eqnarray}
is negative for any finite nonzero value of $r(t)$. The maximum value of $S(\phi,t)$ at $\phi=-\theta(t)$ is 
then given by
\begin{eqnarray}
\nonumber
\max  S(\phi,t) &=& \frac{1}{4} r(t) \cos \Big(\theta(t) + \phi \Big) \Big|_{\phi=-\theta(t)}, \\
&=& \frac{r(t)}{4} = \frac{1}{4} \sqrt{x^2(t) + y^2(t)}.
\label{SMax}
\end{eqnarray}

\begin{figure}[h]
\includegraphics[width=\columnwidth]{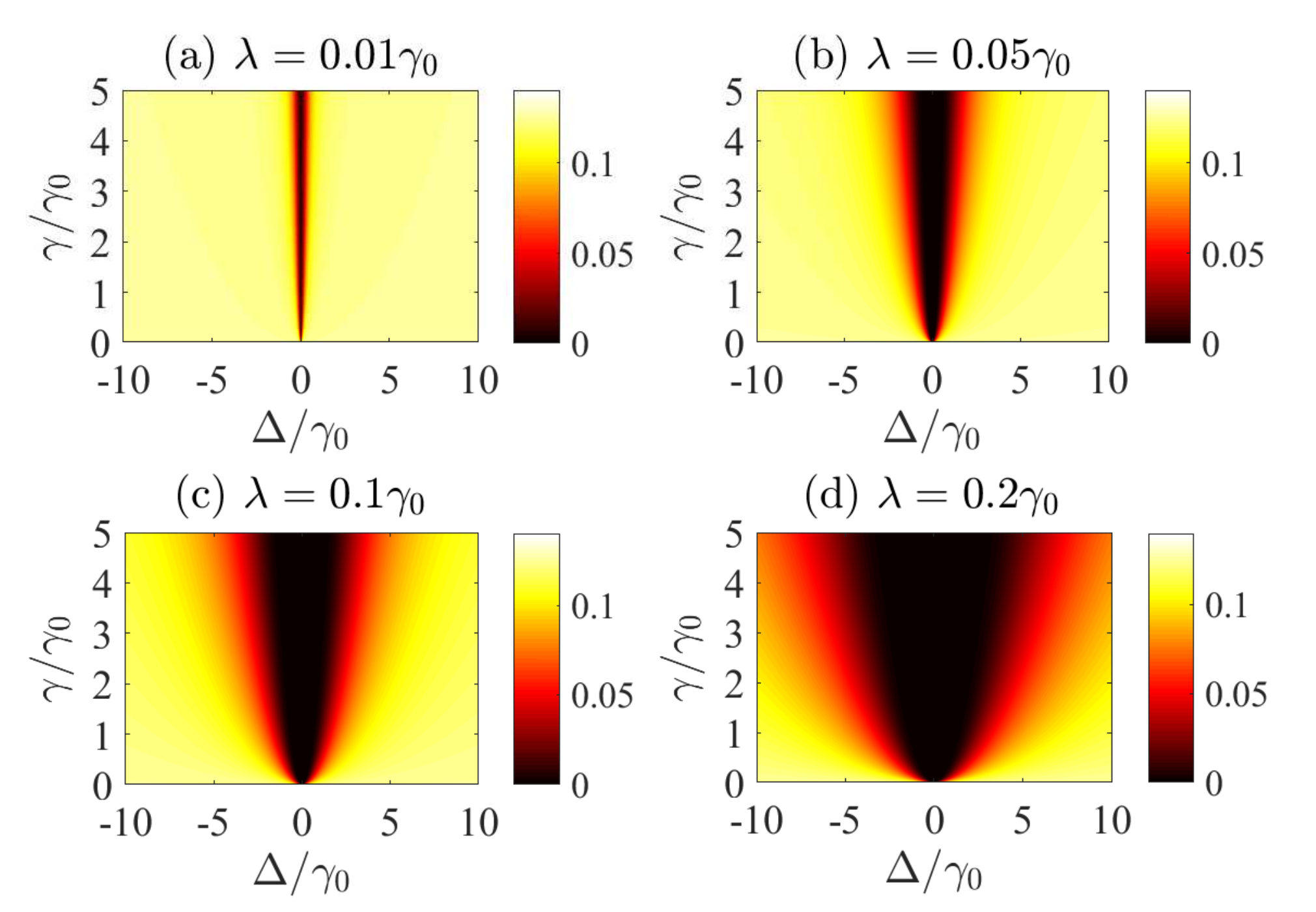}
\caption{This plot describes the maximal value of the shifted phase distribution $S_{m}(t)$ as a function of
the detuning parameter $\Delta$ and coupling strength $\gamma$ for the spectral widths
(a) $\lambda = 0.01\gamma_0$, (b) $\lambda = 0.05\gamma_0$, (c) $\lambda = 0.1\gamma_0$, and
(d) $\lambda = 0.2\gamma_0$. We consider $\gamma_{0} t = 500$ for all the plots.}
\label{sfig3}
\end{figure}

We show here in Fig.~\ref{sfig3} the maximum value of $S(\phi,t)$ as a function of the coupling
strength $\gamma$ and the detuning parameter $\Delta$ with varying spectral width $\lambda$. 
To vary the coupling strength, we consider the spectral density $J(\omega)$ with coupling strength 
$\gamma$ which varies in units of $\gamma_{0}$. In Fig.~\ref{sfig3}, we consider four different
values of the spectral width (a) $\lambda = 0.01\gamma_0$, (b) $\lambda = 0.05\gamma_0$,
(c) $\lambda = 0.1\gamma_0$ and (d) $\lambda = 0.2\gamma_0$. We have taken $\gamma_{0} t =500$ 
for all the plots in Fig.~\ref{sfig3}. We observe the formation of an Arnold tongue which is characteristic
of a synchronized system. But here the explication of the Arnold tongue plot is somewhat different. 
Unlike the situation discussed in the literature 
\cite{koppenhofer2019optimal,parra2020synchronization,roulet2018synchronizing}, 
we find there is no synchronization in the dark tongue region. Rather, the system will have phase localization or 
phase synchronization outside the tongue region. The width of the dark tongue region (unsynchronized region) 
increases with the spectral width $\lambda$.


\end{document}